\documentclass[11pt]{article}
\usepackage{amssymb,latexsym,amsmath}
\usepackage{amsfonts, graphics}
\newtheorem{theorem}{Theorem}

\newcommand{\R}{\mathbb R}

\begin{document}
\title{On the steady states of the spherically symmetric 
Einstein-Vlasov system}
\author{H{\aa}kan Andr\'{e}asson\\
        Department of Mathematics, Chalmers,\\
        S-41296 G\"oteborg, Sweden\\
        email\textup{: \texttt{hand@math.chalmers.se}}\\
        \phantom{tom rad}\\
        Gerhard Rein\\
        Department of Mathematics, University of
        Bayreuth,\\
        D-95440 Bayreuth, Germany\\
        email\textup{: \texttt{gerhard.rein@uni-bayreuth.de}}}

\maketitle

\begin{abstract}
Using both numerical and analytical tools we study
various features of static, spherically symmetric solutions
of the Einstein-Vlasov system. In particular, we investigate
the possible shapes of their mass-energy density and find that they
can be multi-peaked, we give numerical 
evidence and a partial proof for the conjecture that the Buchdahl
inequality $\sup_{r > 0} 2 m(r)/r < 8/9$, $m(r)$ the quasi-local mass, 
holds for all 
such steady states---both isotropic {\em and} anisotropic---,
and we give numerical evidence and a partial proof
for the conjecture that for any given microscopic
equation of state---both isotropic {\em and} anisotropic---the
resulting one-parameter family of static solutions generates a
spiral in the radius-mass diagram.
\end{abstract}

\section{Introduction}
\setcounter{equation}{0}
When studying the properties of non-vacuum spacetimes in general
relativity the choice of the matter model is important.
In the present paper matter is described as a large
ensemble of particles which interact only via the gravitational field 
created by the particles themselves and not via collisions between them.
The distribution of the particles on phase space
is given by a distribution function $f$. This function satisfies the 
Vlasov equation, a first 
order conservation law the characteristics of which are the geodesics
of the spacetime metric, and on the other hand $f$ gives
rise to macroscopic quantities such as mass-energy density, pressure,
and mass current which act as source terms in the Einstein field equations.
For an introduction to
kinetic theory in general relativity and the Einstein-Vlasov system in
particular we refer to \cite{An1} and \cite{Rl7}. 

One distinguishing feature of the Einstein-Vlasov system is that
it allows for a wide variety of static solutions.
The purpose of the present paper is to analyze spherically symmetric 
static solutions both numerically and analytically. 
After some general discussion of such steady states in the next section
and a brief discussion of our numerical approach we investigate 
three different features of the steady states. In Section~\ref{charss}
the energy density as 
a function of the area radius is characterized into three distinct classes. 
In particular, we study steady states 
which as a function of the area radius are supported in an interval
$[R_0,R_1]$ with $R_0>0$,  which we call shells, as well as states 
supported in $[0,R_1]$,  which we call non-shells.
One observation is that anisotropic steady states can have 
a quite rich structure, in particular when compared with their
Newtonian analogues, the steady states of the Vlasov-Poisson system. For 
instance, the energy density can have arbitrarily many peaks and in the case 
of shells these can be separated by vacuum regions so that the steady states 
consist of rings of Vlasov matter. 

Then we consider the question whether there is an upper bound strictly smaller 
than $1$ of the quantity $2M/R_1$,  where $M$ is the ADM mass.
For \textit{isotropic} perfect fluid solutions 
which satisfy the hypothesis that the energy density is non-increasing 
outwards the inequality
\begin{equation} \label{buchdahl}
\frac{2M}{R_1}<\frac{8}{9}, 
\end{equation}
was shown by Buchdahl \cite{Bu}. We prove a pointwise version
of this inequality, cf.\ inequality  (\ref{buchdahl2}) below, 
for a class of steady states of the Einstein-Vlasov
system which includes, but is not restricted to, isotropic states.
Moreover, we give numerical evidence for the conjecture that
all spherically symmetric steady states of the Einstein-Vlasov
system satisfy this inequality, even if the energy density is not
radially non-increasing. 
We mention in passing that the inequality is sharp in the case of shells,
cf.\ \cite{An2}. 

If one prescribes the microscopic equation of state,
i.e., the way in which $f$ depends on the local energy and angular momentum,
one obtains a one parameter family of steady states.
In the last section we investigate the relation of the ADM mass to 
the outer area radius of the steady state along such one-parameter 
families. We prove the existence of mass-radius spirals for
the isotropic case by reducing it to a result of Makino \cite{Ma}
for isotropic fluids, and we give numerical evidence that
these radius-mass spirals are present for any microscopic equation
of state, independently of isotropy, a feature which is again in sharp
contrast to the Newtonian situation. 

\section{Static solutions of the spherically symmetric Einstein-Vlasov system}
\label{general}
\setcounter{equation}{0}
In Schwarzschild coordinates the metric of a static spherically 
symmetric spacetime takes the form
\begin{displaymath}
ds^{2}=-e^{2\mu(r)}dt^{2}+e^{2\lambda(r)}dr^{2}
+r^{2}(d\theta^{2}+\sin^{2}{\theta}d\varphi^{2}),
\end{displaymath}
where $r\geq 0,\,\theta\in [0,\pi],\,\varphi\in [0,2\pi]$.
Asymptotic flatness is expressed by the boundary conditions
\begin{displaymath}
\lim_{r\rightarrow\infty}\lambda(r)=\lim_{r\rightarrow\infty}\mu(r)
=0,
\end{displaymath}
and a regular centre requires
\begin{equation} \label{regcent}  
\lambda(0)=0.
\end{equation}
The static Einstein-Vlasov system is given by the Einstein equations 
\begin{eqnarray}
e^{-2\lambda}(2r\lambda'-1)+1
&=&
8\pi r^2\rho, \label{ee11}\\
e^{-2\lambda}(2r\mu'+1)-1
&=&
8\pi r^2 p, \label{ee22}
\end{eqnarray}
together with the static Vlasov equation which can be written as 
\begin{equation}\label{vlasov}
w \,\partial_{r}f -
\left((1+w^{2}+L/r^{2})\mu' - L/r^3\right)\partial_{w}f=0.
\end{equation}
The variables $w$ and $L$ can be thought of as the momentum in the radial direction 
and the square of the angular momentum respectively.
The matter quantities are given by
\begin{eqnarray}
\rho(r)
&=&
\frac{\pi}{r^{2}}
\int_{-\infty}^{\infty}\int_{0}^{\infty}\sqrt{1+w^{2}+L/r^{2}}
f(r,w,L)\,dL\,dw,
\label{rho} \\
p(r)
&=&
\frac{\pi}{r^{2}}\int_{-\infty}^{\infty}\int_{0}^{\infty}
\frac{w^{2}}{\sqrt{1+w^{2}+L/r^{2}}}f(r,w,L)\,dL\,dw, \label{p}
\end{eqnarray}
where $\rho$ denotes the
mass energy density and $p$ is the radial pressure. 

Before we discuss how solutions of this system can be obtained we
note some additional equations which follow from the above.
Let the quasi-local mass be defined by 
\begin{equation} \label{mdef}
m(r) := 4\pi \int_0^r s^2\rho(s)\, ds. 
\end{equation}
The ADM mass is then $M=\lim_{r\to\infty}m(r)$. Using $m$, 
the field equation (\ref{ee11}) together with the boundary condition
(\ref{regcent})
imply that 
\begin{equation} \label{lambdaeq}
e^{-2\lambda(r)}=1-\frac{2m(r)}{r},
\end{equation}
and by (\ref{ee22}),
\begin{equation} \label{mupeq}
\mu'(r)=e^{2\lambda(r)}\left(\frac{m(r)}{r^2} + 4 \pi r p(r)\right).
\end{equation}
In particular, $\mu$ is radially increasing.
Eqns.~(\ref{ee11}) and (\ref{ee22}) are the $00$ and $11$ components
of the Einstein field equations, and given the matter quantities
they together with the boundary conditions suffice to determine the metric.
But the $33$ component of the Einstein equations is also non-trivial and reads
\begin{equation} \label{ee33}
e^{-2\lambda}\left( \mu'' + (\mu' + 1/r)(\mu' - \lambda')\right) = 8 \pi p_T,
\end{equation}
where the tangential pressure $p_T$ is given by
\begin{equation} \label{pt}
p_T (r) =
\frac{1}{2} \frac{\pi}{r^{4}}\int_{-\infty}^{\infty}\int_{0}^{\infty}
\frac{L}{\sqrt{1+w^{2}+L/r^{2}}}
f(r,w,L)\,dL\,dw.
\end{equation}
Eqn.~(\ref{ee33}) follows from (\ref{ee11})--(\ref{p});
the also non-trivial $44$ component is a multiple of the former.
Next we note that
the Tolman-Oppenheimer-Volkoff equation
\begin{equation} \label{tov}
p' = \frac{2}{r} (p_T - p) - \mu' (\rho + p)
\end{equation}
holds for any sufficiently regular static solution of the spherically
symmetric Einstein-Vlasov equation, as can be seen by a simple 
computation, using (\ref{vlasov}) to express $p'$. 
Finally, if we add  (\ref{ee11}) and (\ref{ee22}) we obtain the 
identity
\begin{equation} \label{lpmp}
\lambda' + \mu' = 4 \pi r e^{2\lambda} (\rho + p).
\end{equation}

Let us now briefly discuss how one can establish the existence
of static solutions of the spherically symmetric Einstein-Vlasov system;
for more details we refer to \cite{R0,R1,RR3,RR4}. Let 
\[
E := e^{\mu(r)} \sqrt{1+w^{2}+L/r^{2}}
\]
denote the local or particle energy which like $L$ is conserved along
characteristics of the static Vlasov equation (\ref{vlasov}). 
The ansatz 
\begin{equation}
f(r,w,L)=\Phi(E,L) \label{ansats}
\end{equation}
satisfies the Vlasov equation, and the macroscopic matter
quantities become functionals of the metric coefficient $\mu$.
Substituting these into Eqn.~(\ref{mupeq}) and  using (\ref{lambdaeq}) 
the above system reduces to a single first order equation for $\mu$.
Analyzing the latter 
constitutes an efficient way to prove the existence of static 
solutions with finite ADM mass and finite extension. 
It should be pointed out that spherically symmetric static solutions which 
do not globally have the form (\ref{ansats}) exist, cf. \cite{Sc2}.
This contrasts the Newtonian case where all spherically symmetric static 
solutions have the form (\ref{ansats}), a fact known as Jeans' Theorem,
cf.~\cite{Ba}. 
As a matter of fact, below we obtain steady states which are good candidates 
for solutions which are not globally of the form (\ref{ansats}). 

In \cite[Thm.~2.1]{RR4} it has been shown that in order to
obtain a steady state of finite ADM mass by the above approach there
must exist a cut-off energy $E_0$ such that 
$\Phi(E,L)=0$ for $E>E_0$ and $L\geq 0$. If $\Phi$ and in particular
such a cut-off energy is prescribed we obtain a steady state solution
by prescribing $\mu(0)$ and solving---numerically or analytically---the
resulting equation (\ref{mupeq}) radially outward.
However, the resulting solution will
in general not satisfy the boundary condition $\mu(\infty)=0$,
but it will have some finite limit $\mu(\infty)$. One can then
shift both the cut-off energy and the solution by this limit
to obtain a solution which satisfies the boundary condition 
$\mu(\infty)=0$. This does not affect $m$ and hence also not $\lambda$,
where it should be noted that the boundary conditions for the latter
both at the centre and at infinity follow from (\ref{lambdaeq}),
provided that the solution has finite ADM mass and that $\rho$
is not too singular at the centre.

As long as one is interested in only a single steady state at a time
this way of handling the boundary condition for $\mu$ at infinity is
acceptable, but it is awkward when one studies families of such steady
states, since $E_0$ and $\mu(0)$ cannot be treated as two free
parameters if one insists on the boundary condition at infinity.
Hence we found it convenient to use an ansatz where 
the resulting equation for $\mu$ can be rewritten in terms
of the function $y(r) = e^{\mu(r)}/E_0$. 
The following is sufficiently general for our purposes:
\begin{equation}\label{ransatz}
f(r,w,L) = \phi(E/E_0) (L-L_0)_+^l,
\end{equation}
where $l>-1/2$, $E_0>0$, $L_0 \geq 0$, $x_{+}:=\max\{x,0\}$ denotes the positive part, 
and  $\phi : ]0,\infty[ \to [0,\infty[$ is measurable, $\phi(\eta)=0$ for $\eta > 1$, 
and there exist constants $k>-1$ and $C>0$ such that 
\[
\phi (\eta) \leq C (1 - \eta)^k,\ \eta \in ]0,1[.
\]
With the ansatz (\ref{ransatz}),
\begin{eqnarray}
\rho (r) 
&=&
c_l  r^{2l} (1+L_0/r^2)^{l+2} \Bigl( g_{l+3/2} +
g_{l+1/2}\Bigr)(\sqrt{1+L_0/r^2} y(r)) , 
\label{rhoyrel}\\ 
p(r)
&=&
\frac{c_l}{2l+3} r^{2l} (1+L_0/r^2)^{l+2} g_{l+3/2}(\sqrt{1+L_0/r^2} y(r)), \label{pyrel}\\
p_T(r)
&=&
(l+1)\, p(r) + 
\frac{c_l}{2} L_0 r^{2l-2} (1+L_0/r^2)^{l+1} g_{l+1/2}(\sqrt{1+L_0/r^2} y(r)), \nonumber
\end{eqnarray}
where for $j>-1,\ l>-1$, and $u\in ]0,1]$,
\begin{equation}
g_j (u)  
:=
\int_1^{1/u}  \phi(u\eta) (\eta^2-1)^j d\eta 
= u^{-(2j+1)} \int_u^1  \phi(\eta) (\eta^2-u^2)^j d\eta,\label{g_j}
\end{equation}
$g_j (u):=0$ for $u>1$, and
\[ 
c_l
:=
2 \pi \int_0^1 s^l (1-s)^{-1/2} ds.
\]
The equation to be solved for $y$ then becomes
\begin{equation} \label{yeq}
y' = \frac{y}{1-\frac{2 m(r,y)}{r}} \left(\frac{m(r,y)}{r^2} + 4 \pi r \,p(r,y)
\right),
\end{equation}
where the dependence of $m$ and $p$ on $y$ arises via
the above formulas (\ref{rhoyrel}) and (\ref{pyrel}).
 
In this way the cut-off energy $E_0$
disappears as a free parameter of the problem, and for $\phi,\ l$, and $L_0$
fixed we obtain a one-parameter family of solutions parameterized by 
$y(0) >0$. Once $y$ is determined on the support of the solution 
its limit $y(\infty)$ at infinity can be computed, and if we define
$E_0:=1/y(\infty)$ and $e^{\mu(r)}:=E_0 y(r)$ we have a steady state 
with the proper boundary condition at infinity.

Notice that if $l=0=L_0$, i.e., the microscopic equation of state
(\ref{ransatz}) does not depend on angular momentum, then $p_T=p$,
i.e., tangential and radial pressure are equal, which is why such steady 
states of the Einstein-Vlasov system are called {\em isotropic}. 

If $j>-1$ and $k+j+1>0$ 
then $g_j \in C(]0,\infty[)$.
If $j>0$ and $k+j>0$ then
$g_j \in C^1(]0,\infty[)$
with 
\begin{equation}\label{hprime}
g'_j (u) = - \frac{1}{u}\left((2 j +1) \, g_j(u) + 2 j g_{j-1} (u)\right),
\end{equation}
cf.\ \cite[Lemma 2.2]{RR4}. With this regularity of the matter terms as
functions of $y$ it is easy to see that Eqn.~(\ref{yeq})
has for any choice of $y(0)>0$ a unique solution. 
The non-trivial question is which choices for $\phi$ lead to
steady states of finite ADM mass and finite extension.
In \cite[Thm.~3.1]{RR4} it has been shown that this is the
case if $L_0=0$ and
\begin{equation} \label{phiass}
\phi(\eta) = c (1 - \eta)^k + \mathrm{O} ((1-\eta)^{k+\delta}) \ \mbox{as}\ \eta \to 1-
\end{equation}
where $c >0$, $\delta >0$, and $k,l \in \R$ are such that
\[
k>-1,\ l>-\frac{1}{2},\ k+l+\frac{1}{2} > 0,\
k<l+\frac{3}{2} .
\]
In this case the resulting steady state is non-trivial iff 
$y(0) \in ]0,1[$.

To conclude this section we briefly discuss how
we construct the steady states numerically. If an ansatz
of the form (\ref{ransatz}) is given, the value of the integral
$g_j(u)$ and hence of $\rho$ and $p$ as functions of $r$ and 
$y(r)$ can easily be computed.
In practice we preferred to choose ansatz functions of the
form 
\begin{equation} \label{pol}
f(r,w,L) = (1-E/E_0)^k_{+}(L-L_0)_{+}^l,
\end{equation}
with $k\geq 0,\ l>-1/2,\ k< 3 l + 7/2$, $E_0>0$, $L_0\geq 0$,
and $k,l$ such that the relevant integrals $g_j$ can be computed
explicitly by hand. In \cite{R1} it has been shown that such an
ansatz leads to finite ADM mass and compact support as well.
In the Newtonian case with $l=L_0=0$,  this ansatz gives
steady states with a polytropic equation of state. 
With $\rho$ and $p$ then given as functions of $r$
and $y(r)$ we use a simple Euler-type or a leap-frog
scheme to solve the equation (\ref{yeq}), starting with some prescribed
value $y(0)\in ]0,1[$ at the centre and moving with a fixed step size
$\Delta r$ radially outward. If we are in a situation where the
rigorous results cited above guarantee that the solution
has a finite extension, the expression
$\sqrt{1+L_0/r^2} y(r)$ exceeds the threshold $1$
for sufficiently large values of $r$. Once this happens
the computation can be stopped, and the spacetime can be extended
by an exterior Schwarzschild solution of the appropriate ADM mass.

It should be noted that when $L_{0}>0$ the expression
$\sqrt{1+L_0/r^2} y(r)$ will always exceed $1$
for sufficiently small values of $r$, and there will be no matter in the region
\begin{equation} \label{r0}
r<\sqrt{\frac{L_0}{y(0)^{-2}-1}}=:R_0. 
\end{equation}
In this case $y$ is constant in the inner vacuum region
$[0,R_0]$, and we start our numerical computation at the radius $R_0$.
The choice $L_0>0$ also leads to another complication,
since in that case the support of the solution may in general
consist of several concentric shells, a feature which is
not captured if the computation is stopped the first time that
$\sqrt{1+L_0/r^2} y(r)$ again exceeds $1$.
 
\section{Characterization of steady states} \label{charss}
\setcounter{equation}{0}
In this section we are interested in the possible shapes of the 
energy density $\rho$ and its support. In particular we will numerically construct
multi-peaked steady states which seem to have no analogue in the Newtonian situation. 

We start with some simple analytic observations. Without explicitly
mentioning it we only consider non-trivial steady states
of finite ADM mass and compact support.
In particular, $m(r)>0$ for some area radius $r$, and hence $m>0$
on an interval of the form $]R_0,\infty[$ where $R_0\geq 0$
is given by (\ref{r0}). Notice that if $L_0>0$ then at $r=R_0$
by definition $y \sqrt{1+L_0/r^2} = 1$, and
$\frac{d}{dr} y \sqrt{1+L_0/r^2} = - y (1+L_0/r^2)^{-1/2} L_0/r^3 < 0$
so that $y \sqrt{1+L_0/r^2} < 1$ on some interval of the form
$]R_0,R_0 + \delta[$. 

\smallskip

\noindent
{\em Observation 3.1}. 
The function $y$ is constant on $[0,R_0]$ and strictly increasing
on $[R_0,\infty[$. This follows immediately from (\ref{yeq}).

\smallskip

\noindent
{\em Observation 3.2}.
If the solution is isotropic, i.e., $l=0=L_0$, then $\rho$ is strictly
decreasing and its support is an interval of the form $[0,R_1]$.

\smallskip

\noindent
{\em Observation 3.3}.
If $L_0=0$, then the support of the solution is still 
an interval of the form $[0,R_1]$. 
This follows from Observation~(3.1), Eqn.~(\ref{rhoyrel}),
and the strict monotonicity of the functions $g_j$ on their support,
cf.~(\ref{hprime}).
Note however that in the anisotropic
case $l\neq 0$ the energy density $\rho$ is in general no longer
monotone.

\smallskip

\noindent
{\em Observation 3.4}.
If $L_0>0$, then we have vacuum in the interval $[0,R_0]$, 
$m(r) >0$ for $r>R_0$, and the support is contained in some interval
$[R_0,R_1]$. As we shall see below the support will in general no
longer be a single interval.

\smallskip

Steady states which are supported in $[R_0,R_1]$ with $R_0>0$ we 
call shells, and states with $R_0=0$ we call non-shells. 
In Section~3.3 below we give a quite general characterization
of both shells and non-shells for different values of the parameters 
$k,l,L_0$ and $y(0)$. 
The main focus in that section is on cases which have rather small values 
on $y(0)$ since this is a necessity for obtaining several peaks. In the 
following sections we have chosen to study shells and non-shells separately 
for values of $y(0)$ which give no more than two peaks. One reason for this 
is that it is then easier to visualize the large difference in magnitude of the first 
peak compared to the following one for these values of $y(0)$; in 
Section~3.3 we really use the quantity $4\pi r^2\rho$ rather than $\rho$ 
itself and moreover we use a logarithmic scale in order to get more 
informative pictures for the multi-peaks. The second reason is that 
we want to emphasize the important feature that shells can have separating 
vacuum regions whereas non-shells cannot, and we find that this point is 
made more clear by first splitting the presentation into shells and non-shells. 

\subsection{Shells}
Since the main purpose of this and the next section is to emphasize a couple of features 
of the shape of the energy density which 
are present for any choice $k\geq 0$ and $l\geq 0$ when $L_0>0$,
we fix $k=0$, $l=1.5$, and $L_0=0.2$. A more general analysis 
where the influence of all parameters is taken into account is carried
out in Section~3.3 below. 

Let us for these values of $k$, $l$, and $L_0$ compute steady states with three different 
values of $y(0)$,  namely $y(0)=0.4,\, 0.22,\, 0.12$. 
\begin{figure}[htbp]
\begin{center}\scalebox{.6}{\includegraphics{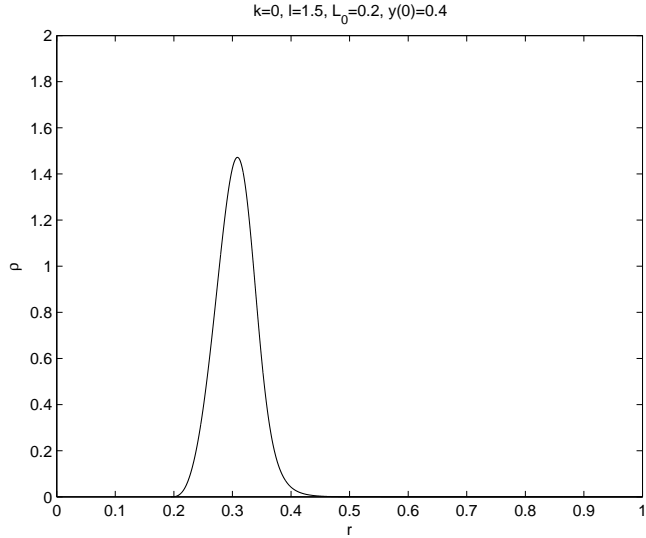}}
\end{center}
\caption{A pure shell}\label{fig1}
\end{figure}
\begin{figure}[htbp]
\begin{center}\scalebox{.6}{\includegraphics{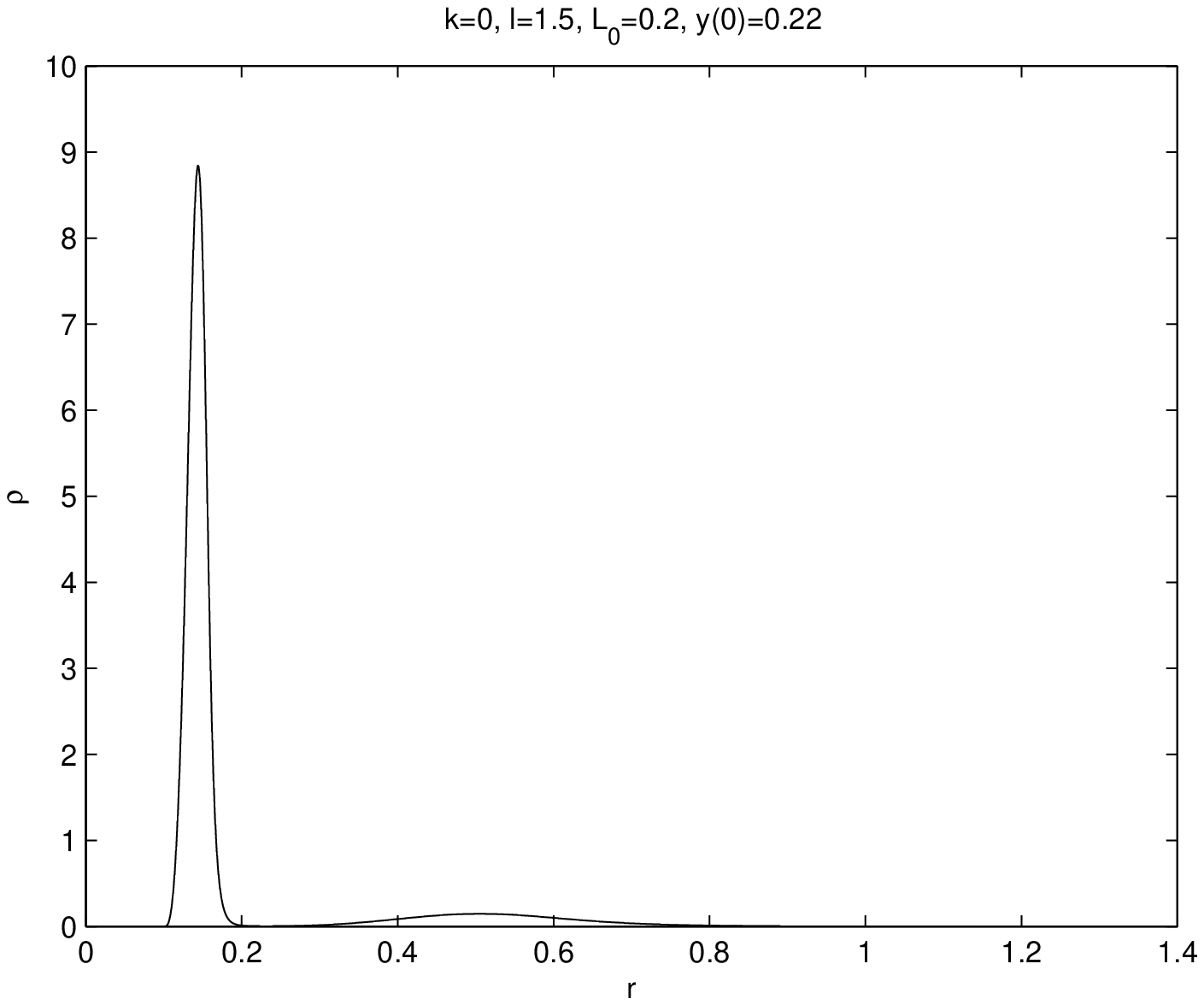}}
\end{center}
\caption{A shell with a tail}\label{fig1}
\end{figure}
\begin{figure}[htbp]
\begin{center}\scalebox{.6}{\includegraphics{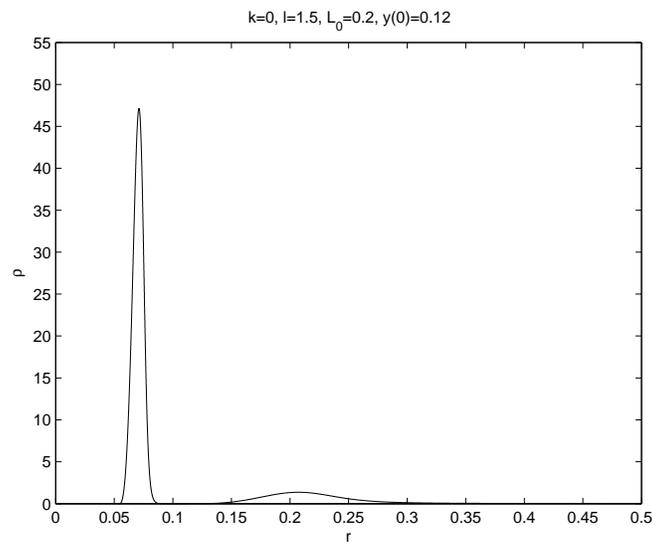}}
\end{center}
\caption{A vacuum region before the tail}\label{fig1}
\end{figure}
We find that these values give rise to three distinct states. 
When $y(0)=0.4 $ we get a pure shell, i.e., the energy density
increases, reaches a maximum value and then decreases to zero and 
remains zero for all larger values on $r$. In the second situation, 
when $y(0)=0.22 $, the maximum value of
$\rho$ is followed by a local minimum, strictly greater than zero (this is hard 
to see in the figure but is easy to check in the data file),
and then a tail with a small amplitude, relative to the maximum
value, but with a considerable extension. In the final
case, where $y(0)=0.12$, there is a vacuum region between the first part, which 
is a pure shell, and the tail. In this case a Schwarzschild solution can 
be joined at the first point where the energy density vanishes 
which results in a different steady state, a pure shell. 

We point out that in \cite{An2} it 
is indeed proved that the energy density vanishes close to $R_0$ if $y(0)$ 
is small which corresponds to the third case above. 
Furthermore, in Section~2 we mentioned that Jeans' Theorem does not 
hold for the Einstein-Vlasov system, cf. \cite{Sc2}, and  
the distribution function $f$ of the steady state obtained by joining
a Schwarzschild solution as described above for $y(0)=0.12$ is not 
globally given as {\em one} function of $E$ and $L$ as in the
ansatz (\ref{ansats}).

\subsection{Non-shells}
In the case when $L_0=0$,  the energy density can be strictly positive or vanish 
at $r=0$ (depending on $l$) but it is always strictly positive sufficiently close 
to $r=0$. Hence, the support of the matter is an interval $[0,R_1]$ with $R_1>0$, 
and we call such states 
non-shells. From Observation 3.3 it follows that no vacuum region can 
separate two parts of a non-shell. 
Except for this fact, the basic features of the shape of the energy density are
preserved for non-shells. In Figures~4 and 5 the energy densities of two non-shells 
are shown, corresponding to $k=0,\; l=1.5,\; y(0)=0.4$ and $k=0,\; l=5.5,\; y(0)=0.1$. We see 
that these have very similar features as the shells in Figures~1 and 2. 
Again we note (cf. Figure~5) 
the large difference in amplitude of the first peak compared to the second one. In view of 
Observation 3.3, which implies that $\rho>0$ in $[0,R_1]$,  it is interesting to note that 
non-shells nevertheless mimic the behaviour of shells close to $r=0$, cf. Figure~5, 
in the sense that the energy density almost vanishes on an interval $[0,R_0]$,  
for some small $R_0>0$. 

Below we will see that the number of peaks is in general not 
restricted to two as in Figures~2, 3, and 5. However, numerically
we found that already 
these solutions with two peaks seem to have no analogue for 
the Vlasov-Poisson system. 
\begin{figure}[htbp]
\begin{center}\scalebox{.6}{\includegraphics{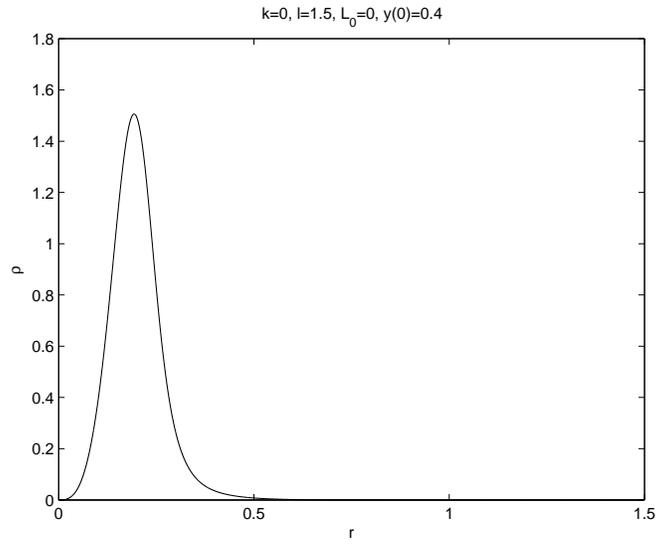}}
\end{center}
\caption{A non-shell without a tail}\label{fig4}
\end{figure}
\begin{figure}[htbp]
\begin{center}\scalebox{.6}{\includegraphics{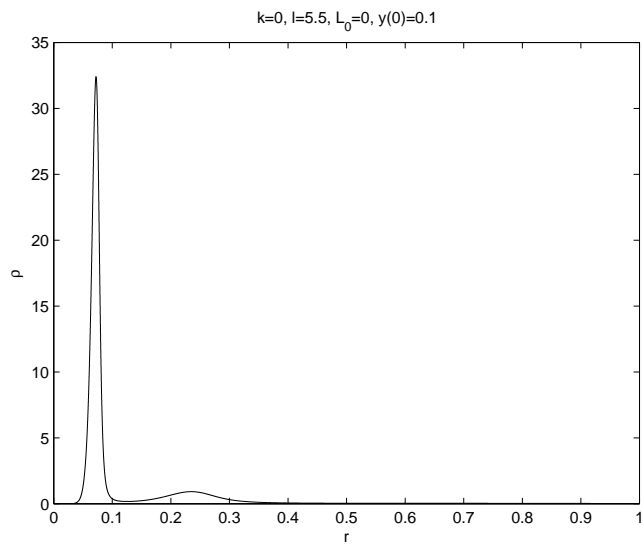}}
\end{center}
\caption{A non-shell with a tail}\label{fig5}
\end{figure}
\subsection{Multi-peaked steady states}
In this section we construct solutions where the energy density has many peaks, and 
we give numerical evidence that there is no limit to the number of the peaks. 
Instead of plotting the energy density $\rho$ itself we plot 
$\tilde{\rho}=4\pi r^2\rho$ (or the logarithm of this quantity) 
since  the pictures become clearer due to the fact that the first peak which 
in amplitude is 
completely dominating gets reduced due to the factor $r^2$. In general the amplitude of 
the peaks get smaller as $r$ increases and thus the quantity $4\pi r^2\rho$ shows less 
difference between the peaks and the resolution in the pictures is improved, but 
we point out that the important features of $\rho$ are similar to those of $\tilde{\rho}$. 

We have first chosen to study the non-shells, i.e., $L_0=0$. In Figure~6 the parameter $k=0$ 
is fixed and the dependence upon $l$ and $y(0)$ is depicted. We see that the parameter 
$y(0)$ has an important effect on the number of the peaks, eg. in the last row
we get from two to four peaks by decreasing $y(0)$ from $0.2$ to $0.05$.  
Increasing the parameter $l$ may also increase the number of peaks as seen
in coloumn 2 and 3. Also the peaks become more 
narrow as $l$ is increased. We point out that more peaks are obtained 
by decreasing $y(0)$ further, cf.\ Figure~11 which shows a shell with $y(0)=0.01$. 

In Figure~7 we have fixed $l=12$ and $y(0)=0.05$ which corresponds to the
final case in Figure~6, and we investigate the influence of $k$. We find 
that by increasing $k$ the decay rate of $\tilde{\rho}$ is affected for 
large $r$ and the peaks in general become wider as $k$ is increased. 

Let us now turn to the shells. We first fix $k=1$, $l=1$, and $L_0=3$ and vary $y(0)$. 
The result is shown in Figure~8. It is instructive to compare this result with the upper 
row in Figure~6. We see that the influence of a non-vanishing $L_0$ increases the amplitude of 
$\tilde{\rho}$ and also more peaks are obtained. The results obtained by varying $L_0$ are
then shown in Figure~9. First of all we notice that the number of peaks is affected, and 
we also note 
that the peaks get more narrow as $L_0$ increases. In particular we 
get in the case when $L_0=1.5$ one vacuum region which separates the first two peaks and 
in the case when $L_0=10$,  there are two vacuum regions. 

The dependence on $l$ is shown in Figure~10 and a similar behaviour as in Figure~6 is 
found. Increasing $l$ can give rise to new peaks and quite generally it makes the peaks 
more narrow. 

Finally we show in Figure~11 a case where $k=1$, $l=10$, $L_0=3$, and $y(0)=0.01$,  which has ten 
peaks, and we note that there are three vacuum regions separating the first four peaks. 







\begin{figure}[htbp]
\begin{center}{\includegraphics{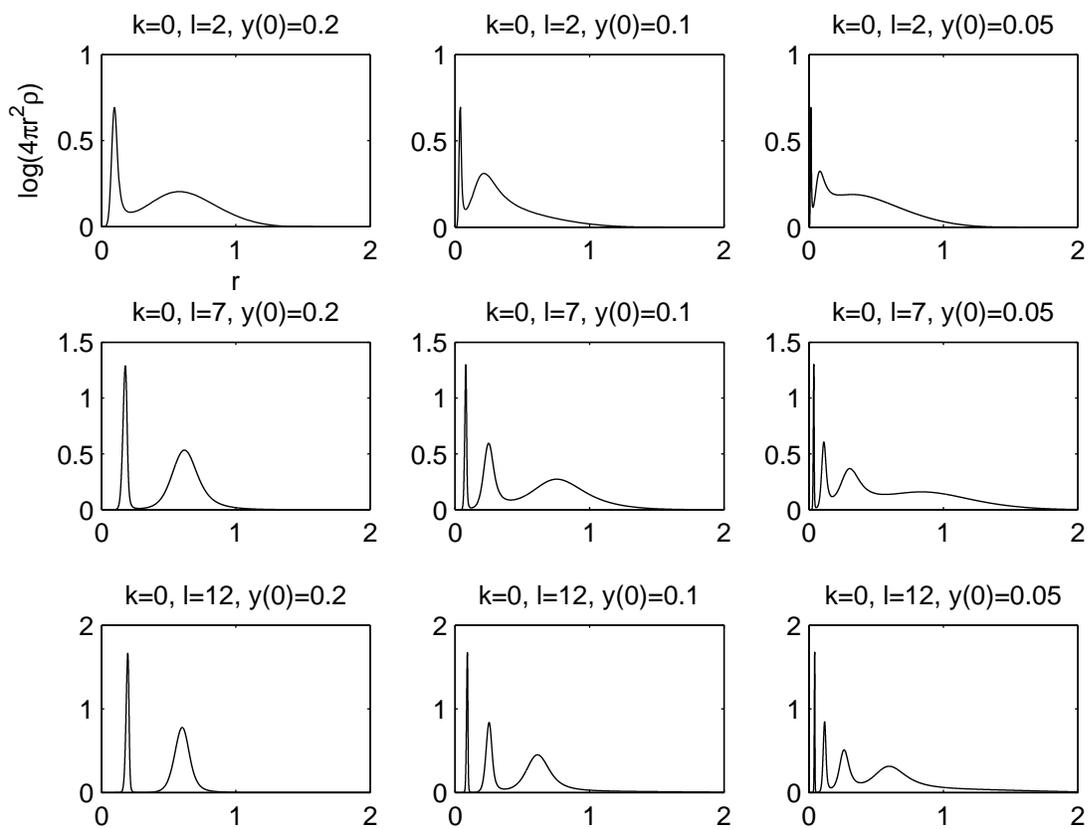}}
\end{center}
\caption{Multi-peaks of non-shells, $L_0=0$}\label{fig6}
\end{figure}
\begin{figure}[htbp]
\begin{center}\scalebox{.8}{\includegraphics{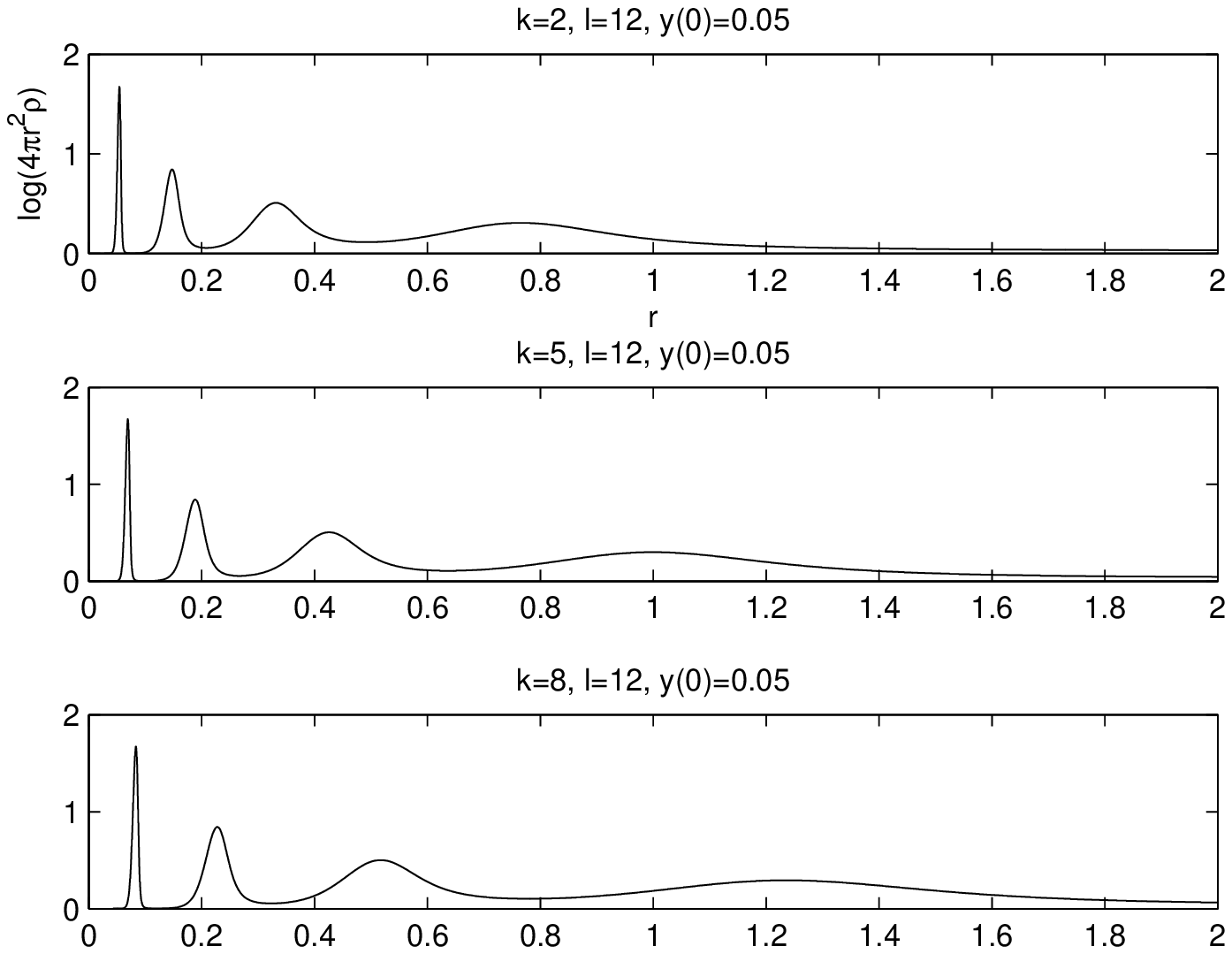}}
\end{center}
\caption{Multi-peaks of non-shells, $L_0=0$}\label{fig7}
\end{figure}
\begin{figure}[htbp]
\begin{center}\scalebox{.8}{\includegraphics{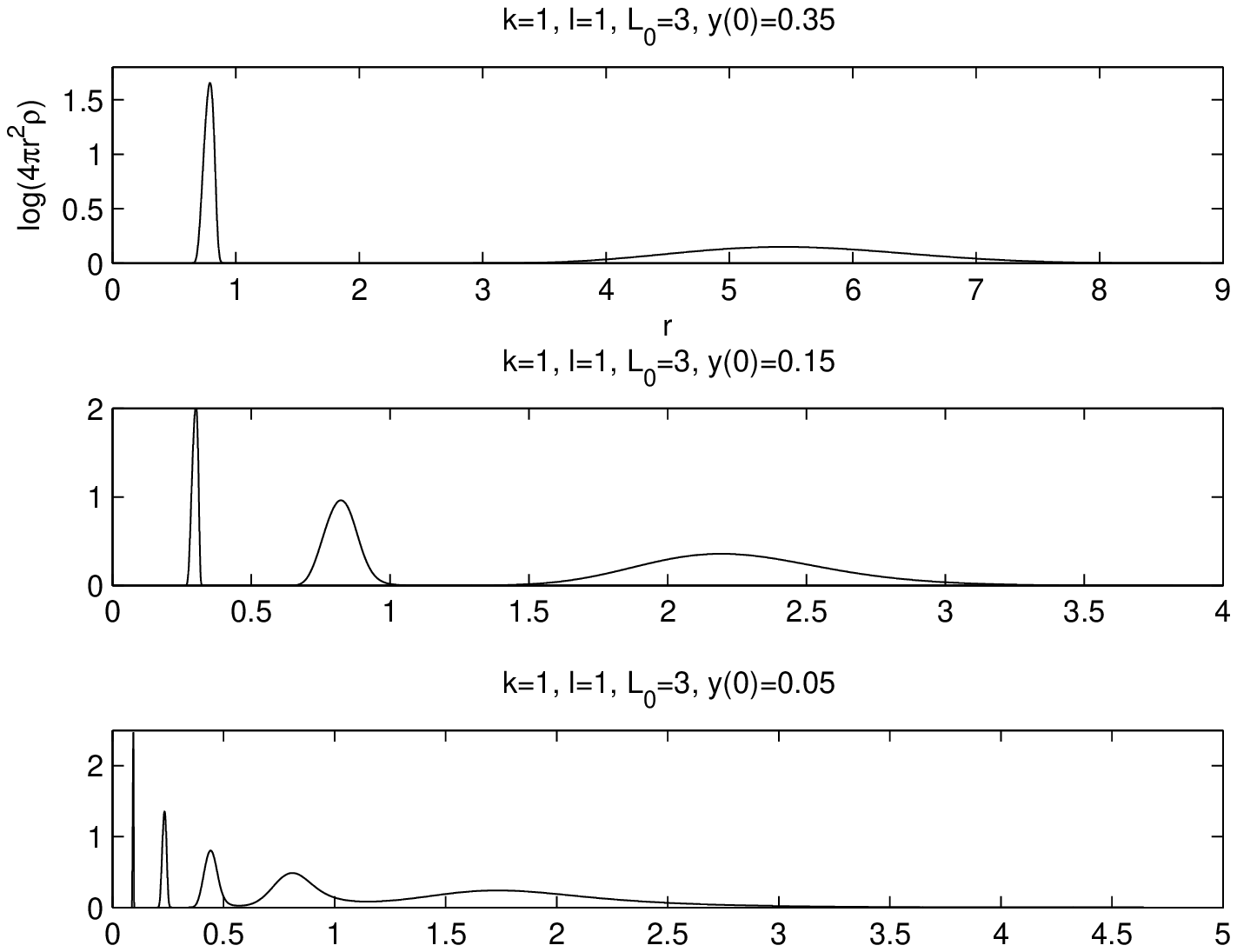}}
\end{center}
\caption{Multi-peaks of shells}\label{fig8}
\end{figure}
\begin{figure}[htbp]
\begin{center}\scalebox{.8}{\includegraphics{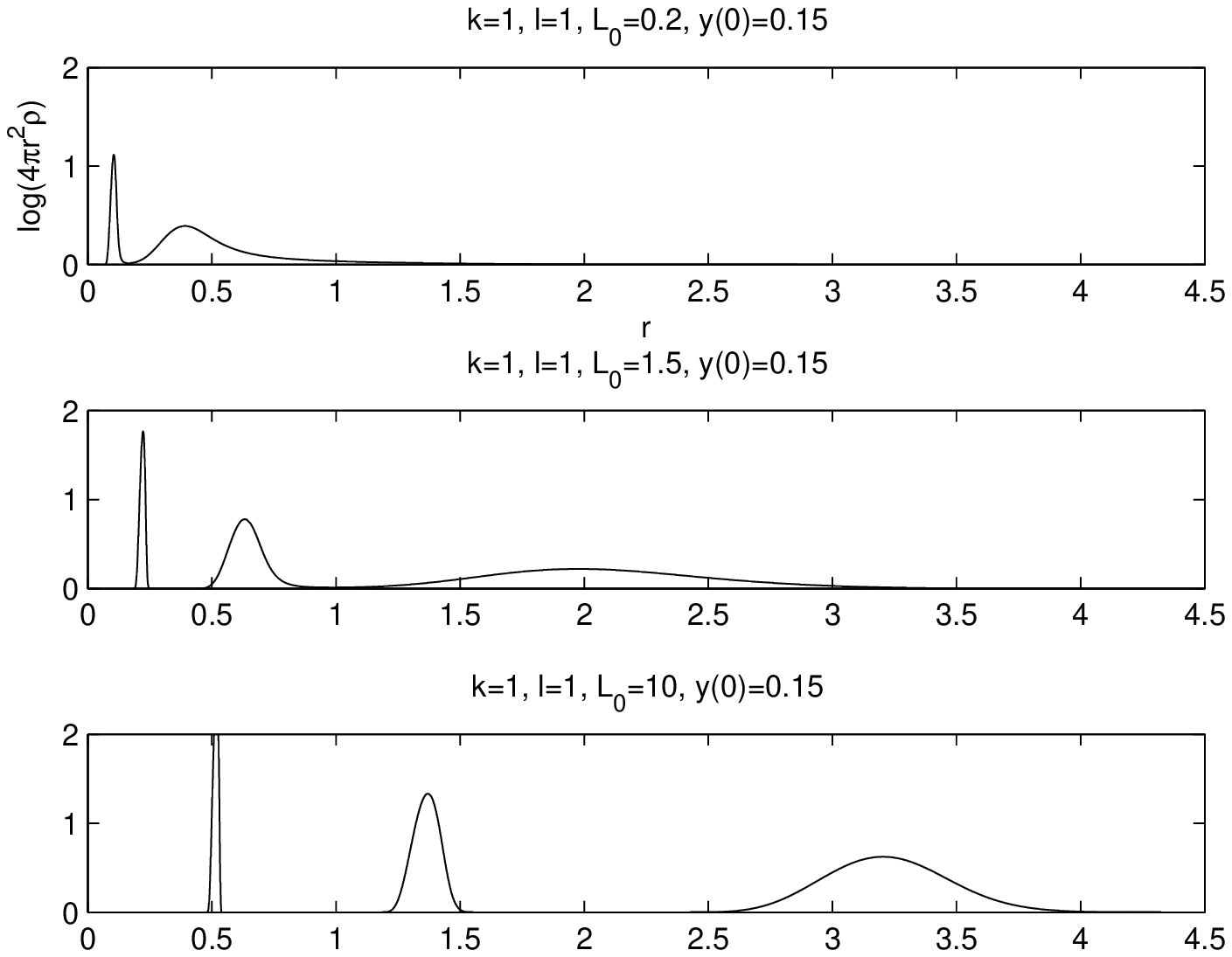}}
\end{center}
\caption{Multi-peaks of shells}\label{fig9}
\end{figure}
\begin{figure}[htbp]
\begin{center}\scalebox{.8}{\includegraphics{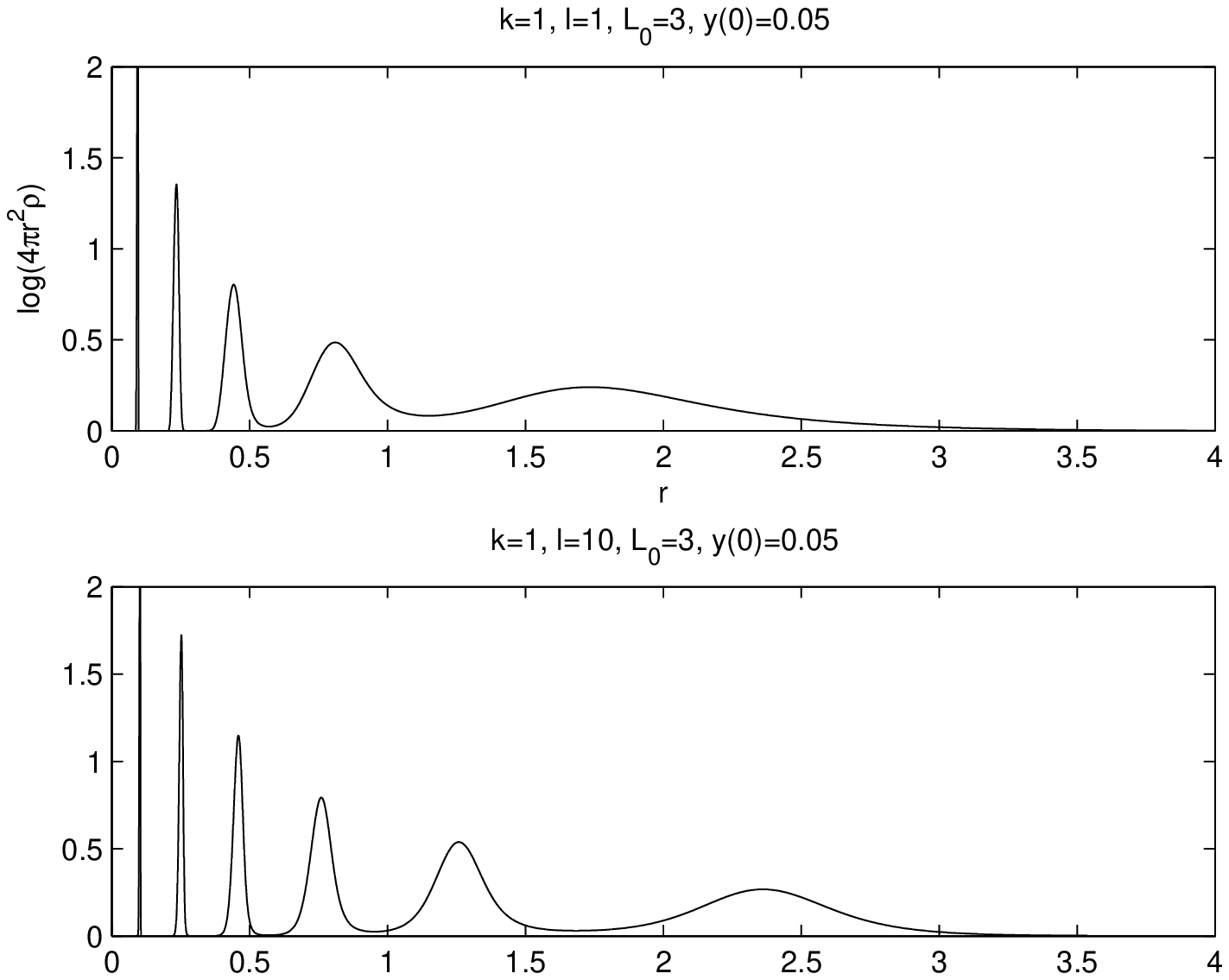}}
\end{center}
\caption{Multi-peaks of shells}\label{fig10}
\end{figure}
\begin{figure}[htbp]
\begin{center}\scalebox{.8}{\includegraphics{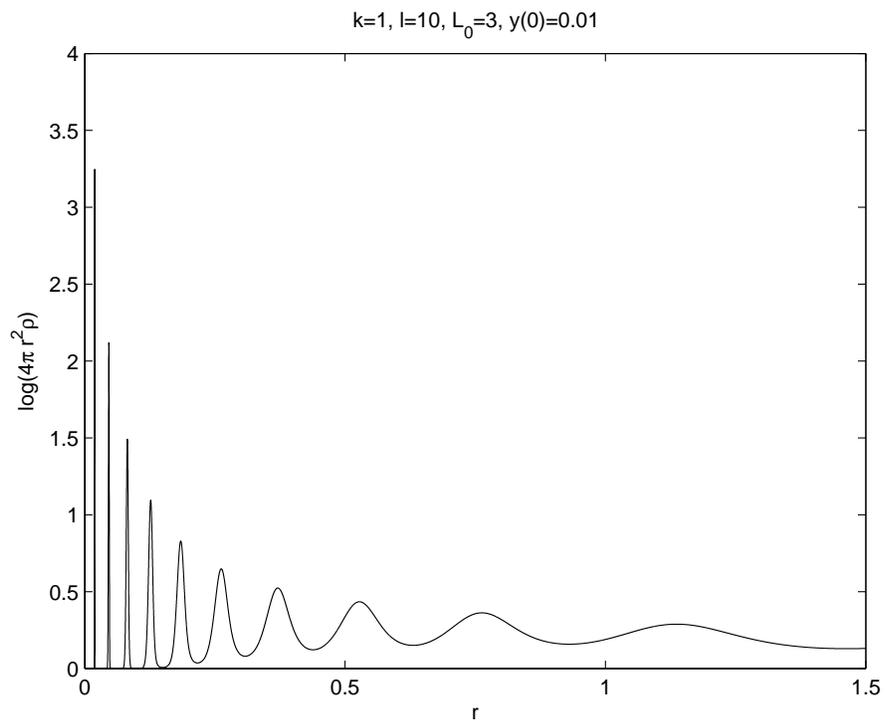}}
\end{center}
\caption{Multi-peaks of a shell}\label{fig11}
\end{figure}
\section{The Buchdahl inequality}
\setcounter{equation}{0}
Let us define
\begin{equation}
\Gamma:=\sup_{r\geq 0}\frac{2m(r)}{r}.\label{Gamma}
\end{equation}
In this section we investigate if $\Gamma$ has
an upper bound less than one for static solutions of the spherically symmetric
Einstein-Vlasov system. 
We start with the following analytic result the proof of which is modeled on
the original proof by Buchdahl.
\begin{theorem} \label{buchdahl}
The Buchdahl inequality
\[
\Gamma < \frac{8}{9}
\]
holds for spherically symmetric steady states of the Einstein-Vlasov
system which satisfy the condition that
\begin{equation}
p \geq p_T \  \mbox{and $\rho$ is non-increasing}.\label{buchass}
\end{equation}
These assumptions are satisfied if $l=0=L_0$ in the ansatz (\ref{ransatz}), 
i.e., the steady state is isotropic, but also if
$L_0=0$ and $l<0$ in which case the steady state is anisotropic.
\end{theorem}
{\bf Proof.}
First of all we notice that using (\ref{lpmp}) the field equation
(\ref{ee33}) can be rewritten in the form
\[
\frac{d}{dr} \left(r^2 e^{\mu-\lambda} \mu'\right)
= 4 \pi r^2 e^{\mu + \lambda} (\rho + p + 2 p_T).
\]
This equation is essentially the slicing condition
for maximal areal coordinates which in the static case coincide
with Schwarzschild coordinates, cf.~\cite{AnRe1}.
Using (\ref{mupeq}) the above equation can be rewritten in the form
\begin{equation} \label{buchbasic}
\frac{d}{dr} \left(\frac{1}{r} e^{\mu-\lambda} \mu'\right)
=
\frac{4\pi}{r} e^{\mu + \lambda}
\left( \rho + 2 p_T - \frac{3}{4\pi} \frac{m}{r^3} - 2 p \right) .
\end{equation}
In the sequel we abbreviate $z(r)=m(r)/r^3$. Since $\rho$ is non-increasing,
\[
z(r) = \frac{4\pi}{r^3}  \int_0^r \rho(s)s^2 ds \geq 
\frac{4\pi}{r^3}  \rho(r) \int_0^r s^2 ds
= \frac{4\pi}{3} \rho(r).
\]
Since by assumption $p_T \leq p$, the function
\[
\frac{1}{r} e^{\mu-\lambda} \mu'
\]
is non-increasing. Since
\[
z'(r) = -\frac{3}{r} z(r) + \frac{4\pi}{r} \rho(r) \leq 0,
\]
the same is true for $z$. We now use 
\[
\xi = \xi(r) = 2\int_0^r s e^{\lambda(s)}ds,\ r\geq 0,
\]
as a new independent variable and consider the function
\[
\alpha(\xi) = e^{\mu(r)}.
\]
Then
\[
\frac{d\alpha}{d\xi} = e^{\mu(r)} \mu'(r) \frac{1}{2 r e^{\lambda(r)}}
\]
is non-increasing for $r > 0$ respectively $\xi > 0$. If we observe that
Eqn.\ (\ref{lambdaeq}) can be rewritten as
\[
e^{\lambda (r)} = (1-2 r^2 z(r))^{-1/2}
\]
then for $r>0$,
\begin{eqnarray*}
\alpha(\xi(r)) 
&>&
\alpha(\xi) - \alpha(0) = \int_0^\xi \frac{d\alpha}{d\xi}(\zeta)\, d\zeta\\
&\geq&
\frac{d\alpha}{d\xi}(\xi)\xi 
= \frac{d\alpha}{d\xi}(\xi) \int_0^r 2\,s\, (1-2 s^2 z(s))^{-1/2} ds\\
&\geq&
\frac{d\alpha}{d\xi}(\xi) \int_0^r 2\,s\, (1-2 s^2 z(r))^{-1/2} ds\\
&=&
e^{\mu(r)} \mu'(r) \frac{1}{2 r e^{\lambda(r)}}
\frac{1}{z(r)}\left(1-\sqrt{1-2 r^2 z(r)}\right).
\end{eqnarray*}
By (\ref{mupeq}), $\mu'(r) \geq e^{2\lambda} r z(r)$,
and inserting this into the previous estimate we find that
\[
e^{\mu(r)} = \alpha(\xi(r)) > \frac{1}{2} e^{\mu(r) + \lambda(r)}
(1-e^{-\lambda(r)}).
\]
This implies that $e^{\lambda(r)} < 3$ which is equivalent to the assertion
on $\Gamma$. 

For the ansatz (\ref{ransatz}) with $L_0=0$ and $l\leq 0$ the 
general assumptions used above are satisfied, and the proof is
complete. \hspace*{\fill} $\Box$  

\smallskip

\noindent
{\bf Remark.} (a)
The analysis above actually yields the somewhat
sharper estimate that 
\[
e^{\mu(r)} - e^{\mu(0)} \geq \frac{1}{2} e^{\mu(r)}(e^{\lambda(r)}-1),
\]
and hence
\[
\Gamma \leq 1-\frac{1}{(3-2e^{\mu(0)})^2}.
\]
(b)
Although our numerical investigation indicates that the
estimate $\Gamma < 8/9$ holds for steady states of the Einstein-Vlasov
system in general, some key features used in the above proof fail, 
in particular, the right hand side of (\ref{buchbasic})
is not negative in general. To see this, let
\[
\tilde \rho (r):= r^{-2l} \rho(r),\ \tilde p (r):= r^{-2l} p(r).
\]
Then if $l>0$ and $L_0=0$,
\begin{eqnarray*}
r^{-2l} \left( \rho + 2 p_T - \frac{3}{4 \pi} \frac{m}{r^3} -2 p \right)
&=&
\tilde \rho + 2 l \tilde p - \frac{3}{4 \pi} \frac{m}{r^{3+2l}}\\
&\to&
\tilde \rho(0) + 2 l \tilde p(0) -  \frac{3}{3+2l} \tilde \rho(0) > 0
\end{eqnarray*}
as $r\to 0$, and the right hand side of (\ref{buchbasic})
is positive for $r$ close to 0.

\smallskip

\noindent
(c)
In \cite{GM} bounds on $\Gamma$ are considered under various assumptions. 
In particular the authors obtain the bound $8/9$ under the assumptions (\ref{buchass}). 
We have chosen to include the above proof anyway for the sake of completeness 
and in order to stress the fact that for Vlasov matter these assumptions are 
satisfied for the ansatz (\ref{ransatz}) with $l\leq 0$ and $L_0=0$ so that steady 
states with these properties do exist. 
In \cite{GM} the authors also consider the general non-isotropic 
case without any monotonicity assumptions but 
instead they require that there is a constant $B$ such that 
\[
\frac{(\rho-3m/r^3)+2(p_T-p)}{p+m/r^3}\leq B.
\]
Under this assumption they obtain a bound on $\Gamma$ which for large $B$ can 
be written in the form
\[
\Gamma\leq 1-2/(2+B)^2.
\] 
As $B\to\infty$ this bound degenerates to
the estimate $\Gamma\leq 1$.  
This is interesting in view of the numerical results presented below 
which indicate that $\Gamma$ is {\em always} less than $8/9$, while
the energy density along a family of steady states 
which makes $\Gamma$ approach $8/9$ gets more and more peaked so that $B\to\infty$ 
(this is easily seen by considering an approximation of a Dirac measure). 
In the case of shells such a family has indeed been constructed 
for Vlasov matter and it has been shown that $\Gamma\to 8/9$ in that case, 
cf. \cite{An2}. We also refer to \cite{An3} where any matter model which satisfies 
$p+2p_T\leq\Omega\rho$ is considered ($\Omega=1$ for Vlasov matter) and where it 
is shown that a family of shells for which $B$ can become arbitrary large satisfies 
$\Gamma\leq (2\Omega+1)^2-1)/(2\Omega)^2$ so that $\Gamma$ is strictly bounded away 
from one. 
Thus, as is shown in (b) above a proof of our general conjecture below 
must rely on mechanisms which are rather different from the ones
in the proof of Theorem~\ref{buchdahl} and also different 
from \cite{GM}. 

\smallskip

Our next aim is to give numerical support to the conjecture
that $\Gamma < 8/9$ holds for all
spherically symmetric static solutions of the Einstein-Vlasov system.
Our numerical study is restricted to
distribution functions of the form (\ref{pol}). Since the qualitative 
behaviour is quite different in the case of shells and 
non-shells we will present the results for these cases separately.
\subsection{Shells}
In Figure~$12$ the values of $\Gamma$ are shown for a family of steady 
states parameterized by $y(0)\in [0.03,0.6]$. Here $k=0$, $l=3/2$, and $L_0=1$. 
We see that $\Gamma$ stays below $8/9$. In Figure~13 we have changed the range of 
$y(0)$,  namely $y(0)\in [e^{-14},e^{-4}]$, in order to see that the family does 
approach $8/9$ as $y(0)\to 0$,  which we know is the case by the result in \cite{An2}. 
This also provides a fairly tough test for our numerical scheme since
the functions $g_j$ become very large if $y$ gets very close to $0$. 
\begin{figure}[htbp]
\begin{center}\scalebox{.6}{\includegraphics{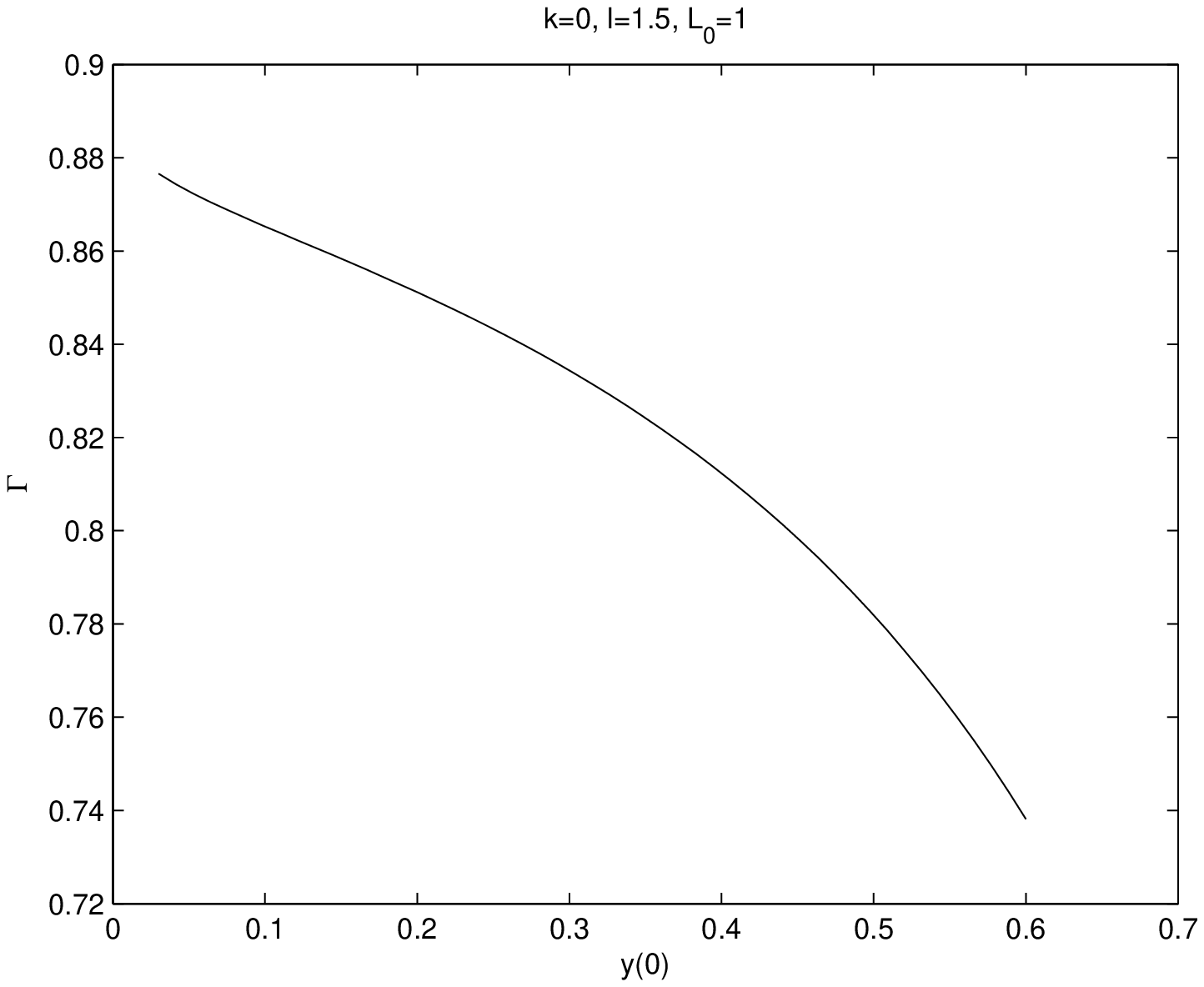}}
\end{center}
\caption{$\Gamma$ versus $y(0)$ for shells}\label{fig8}
\end{figure}
\begin{figure}[htbp]
\begin{center}\scalebox{.6}{\includegraphics{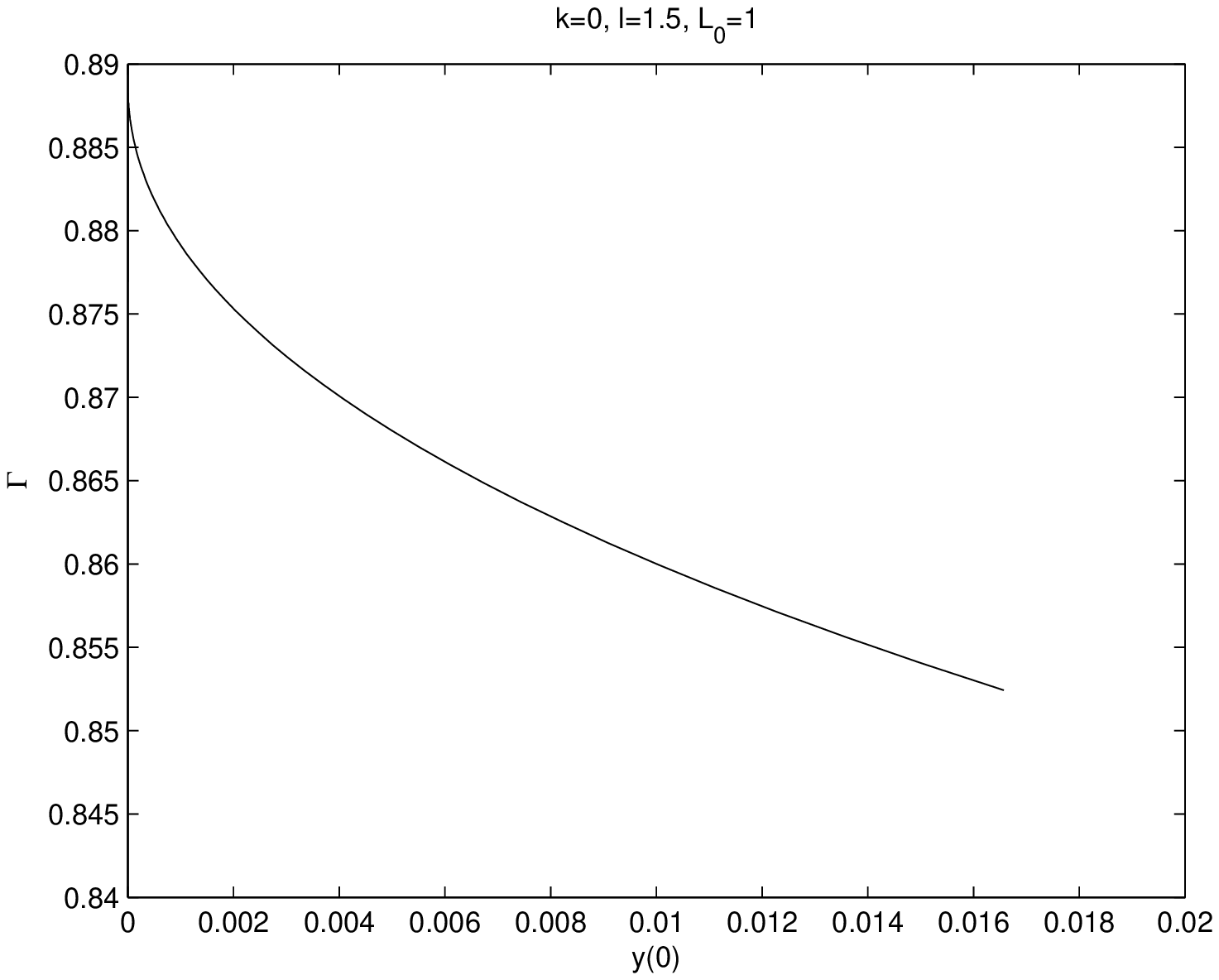}}
\end{center}
\caption{$\Gamma$ versus $y(0)$ for shells}\label{fig8}
\end{figure}

We point out that the corresponding values of $2M/R_1$,  where $R_1$ is 
the outer radius 
of the support would (as long as several peaks are present) be much less than $\Gamma$. 
This is clear in view of Section~3 where it was seen that the first peak is strongly 
dominating compared to the other peaks and most of the matter is captured in the first peak. 
Hence the value of $r$ such that $\Gamma=2m(r)/r$ is within that peak. 

Let $R_{11}$ be the outer radius of the \textit{first} peak and consider 
the ratio $R_{11}/R_0$ as $y(0)\to 0$. We find in Figure~14 that 
$R_{11}/R_0\to 1$ as $y(0)\to 0$. This fact is indeed proved in \cite{An2} and 
is crucial in order to show that $\Gamma\to 8/9$ for any family of 
shells for which $y(0)\to 0$. 
\begin{figure}[htbp]
\begin{center}\scalebox{.6}{\includegraphics{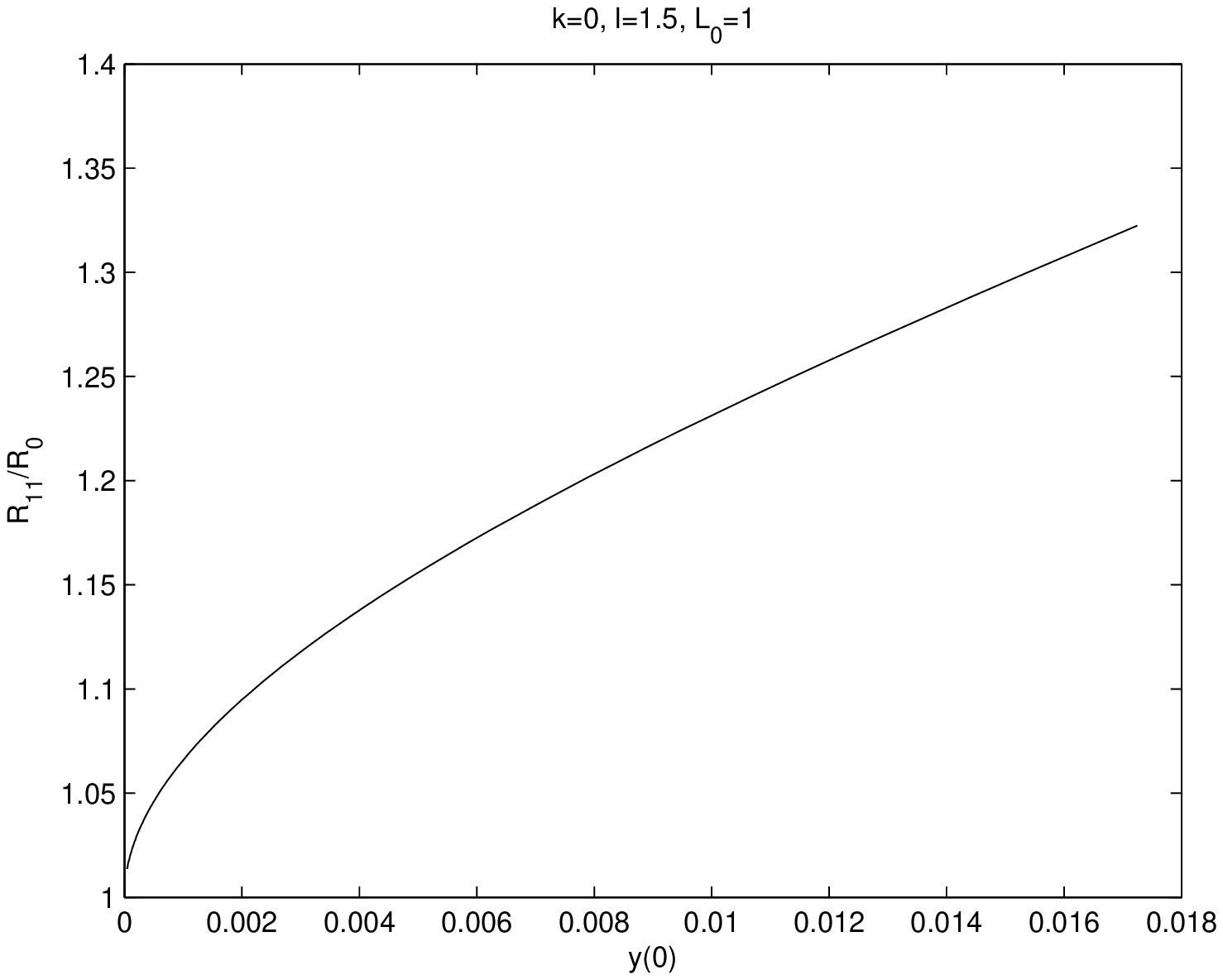}}
\end{center}
\caption{The ratio $R_{11}/R_0$}\label{fig5}
\end{figure}

In contrast to the non-shell case the behaviour of $\Gamma$ for shells 
is qualitatively very similar for different values of 
$k$ and $l$ and we have therefore only included one case here. However we 
point out that all static shells that we have considered have satisfied 
$\Gamma<8/9$.

\subsection{Non-shells}
\begin{figure}[htbp] 
\begin{center}\scalebox{.6}{\includegraphics{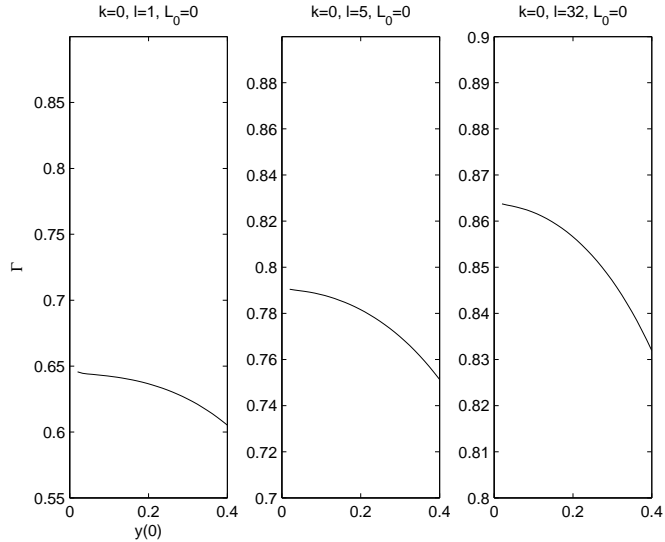}}
\end{center}
\caption{$\Gamma$ versus $y(0)$ for non-shells}\label{gammathreecases}
\end{figure}

In Figure~\ref{gammathreecases} we depict $\Gamma$ as a function of $y(0)$ 
for three different families 
of non-shells, where $l=1, 5, 32$, and $k=0$ in all cases. 
The results indicate that 
\begin{equation}
\Gamma<\kappa(l)<8/9,\label{claim2}
\end{equation}
where $\kappa$ is an increasing function of $l$. 
It is interesting to note the strong dependence of $\Gamma$ on $l$  which shows that 
the degree of anisotropy is crucial. In particular 
we see that the (non-rigorous) results by Bondi in \cite{Bo} can be violated 
by anisotropic steady states. Indeed, Bondi considers isotropic matter models 
for which $\rho\geq 0,\;\rho\geq p$,  or $\rho\geq 3p$,  and gets $0.97,\; 0.86$ 
and $0.70$ respectively as upper bounds of $2M/R_1$. Since for Vlasov matter
$\rho>p$,  we see that the second bound is violated when $l$ is 
large, i.e., the degree of non-isotropy is large, cf. the third case in 
Figure~\ref{gammathreecases}. 
For shells this bound is violated also when $l$ is small as we saw in Figure~12. 
The degree of anisotropy in this case is large for any $l\geq 0$,  
since $p/\rho\to 0$ as $y(0)\to 0$,  cf \cite{An2}. 

By taking $l$ larger we get numerical evidence that also non-shells can have $\Gamma$ 
arbitrary close to  $8/9$,  i.e., $\kappa(l)\to 8/9, \mbox{ as }l\to\infty$,  but 
still $\kappa(l)<8/9$ for any $l$. 
In view of this claim and the claim made in the previous paragraph for shells, 
we find that the most important insight from our simulations in connection 
to Buchdahl's inequality is the following claim: 

Any static solution of the Einstein-Vlasov system satisfies 
\begin{equation}
\Gamma<\frac{8}{9}.\label{buchdahl2}
\end{equation}

As mentioned above, the result in \cite{An2} shows that the inequality is sharp for shells 
and our numerical 
simulations indicate that it is also sharp for non-shells. It is an interesting and important 
problem to prove (\ref{buchdahl2}) in full generality.

%
%
%
\section{Spirals in the $(R,M)$ diagram}
\begin{figure}[htbp]
\begin{center}\scalebox{.6}{\includegraphics{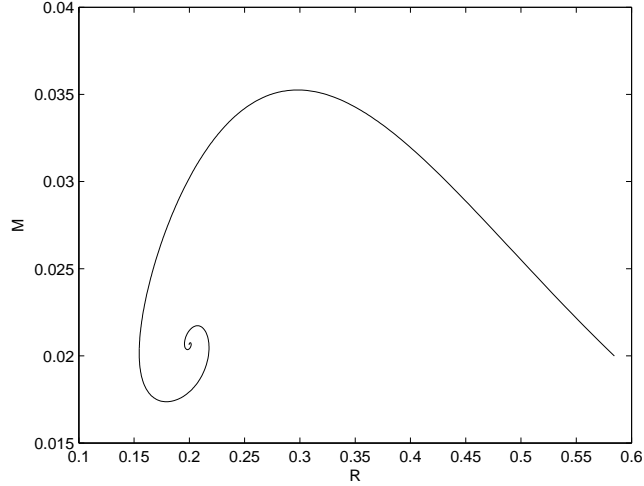}}
\end{center}
\caption{$(R,M)$ spiral for $k=0,\ l=0,\ L_0=0,\ 0.01 \leq y(0) \leq 0.95$}
\label{isospiral}
\end{figure}

\setcounter{equation}{0}
As we discussed in Section~\ref{general} for a fixed
ansatz of the form (\ref{ransatz}) there exists a
one-parameter family of corresponding non-trivial static
solutions which are parameterized by $y(0)\in ]0,1[$ where $y=e^\mu /E_0$.
One can now ask how for example the ADM mass $M$ and the radius 
of the support $R$ change along such a family. More specifically,
one can plot for each $y(0)$ the resulting values for 
$R$ and $M$ to obtain a curve which reflects how radius
and mass are related along such a one-parameter family of steady states.
As a first example we choose $l=0=L_0$ and $k=0$ in (\ref{pol}),
i.e., an isotropic case,
which results in the curve shown in Figure~\ref{isospiral}.
We will discuss possible conclusions from such curves in more detail
below, but note that there are possibly many steady states (or none at all)
which have a given ADM mass and differ in radius.

As a matter of fact, for the isotropic case one can show that such
radius-mass curves always have a spiral form. The following
theorem easily follows from the corresponding result established
by Makino for static, spherically symmetric solutions of the 
Einstein-Euler system under suitable assumptions on the macroscopic
equation of state $p=p(\rho)$, cf.~\cite[Thm.~1]{Ma}.
For this theorem we parameterize our one-parameter family of steady states
via the inverse of the central pressure, i.e.,
we define $\epsilon := \frac{3}{c_0 g_{3/2}(y(0))}$; notice that 
$g_{3/2}$ is one-to-one on $]0,1]$ and $\epsilon \to 0$ iff $y(0) \to 0$. 

\begin{theorem} \label{rmspirals}
Consider an isotropic ansatz of the form (\ref{ransatz}), i.e.,
$l=0=L_0$, and assume that the conditions on $\phi$
specified in (\ref{phiass}) hold. Then as $\epsilon \to 0$,
\[
\left(\begin{array}{r}
R(\epsilon) \\ M(\epsilon)
\end{array} \right)
= \left(\begin{array}{r}
R_S \\ M_S 
\end{array} \right) + 
\epsilon^{\gamma_1} B J(\gamma_2 \ln \epsilon)\, b
+\mathrm{o}(\epsilon^{\gamma_1}).
\]
Here $R(\epsilon)$ and $M(\epsilon)$ denote the radius of the support and the
ADM mass of the steady state corresponding to the parameter $\epsilon$,
$R_S$ and $M_S$ are positive constants, $B$ is a non-singular matrix,
\[
J(\theta):= 
\left(
\begin{array}{rr}
\cos \theta & -\sin \theta \\ 
\sin \theta & \cos \theta
\end{array} \right)
\]
is a rotation by the angle $-\theta$, $b\in \R^2\setminus \{0\}$,
and $\gamma_1,\ \gamma_2$ are positive constants.
\end{theorem}

\noindent
{\bf Proof.}
The definition of $m$ in (\ref{mdef}) together with the
Tolman-Oppenheimer-Volkoff equation (\ref{tov}) in the isotropic case
with (\ref{mupeq}) and (\ref{lambdaeq}) substituted in give the following
system of equations:
\begin{eqnarray}
\frac{dm}{dr} 
&=&
4 \pi r^2 \rho,\label{tov1}\\
\frac{dp}{dr} 
&=&
-(\rho + p) \frac{m + 4 \pi r^3 p}{r^2(1-2m/r)}. \label{tov2}
\end{eqnarray}
Moreover, 
\[
\rho = c_0 \left( g_{3/2}(y) + g_{1/2}(y) \right),
\]
where the right hand side is a strictly decreasing function
of $y\in ]0,1]$ which converges to infinity as $y\to 0$.
If we denote its inverse by
\[
\sigma : [0,\infty[ \to ]0,1],\ \rho \mapsto y,
\]
we find that the pressure $p$ is related to the energy density $\rho$
through an equation of state
\begin{equation} \label{eqofstate}
p = P(\rho) := \frac{c_0}{3} g_{3/2} (\sigma(\rho)).
\end{equation}
The equations (\ref{tov1}), (\ref{tov2}), (\ref{eqofstate})
now provide a closed system of equations which describes 
an isotropic Einstein-Vlasov steady state purely in terms
of macroscopic quantities and a macroscopic equation of state.
This system also arises from the static, spherically symmetric
Einstein-Euler system. It describes a gaseous star and is studied
in \cite{Ma}. Hence to prove the theorem above we only have to
show that our equation of state (\ref{eqofstate}) satisfies
the assumptions made in \cite{Ma}, and then we can invoke 
the corresponding result established there.

Firstly, $P=P(\rho)$ is a smooth function of $\rho>0$,
since $\rho>0$ iff $y\in ]0,1[$ iff $p>0$ we have 
$P(\rho) > 0$ for $\rho>0$ and since 
$\rho\to 0$ iff $y\to 1$ iff $p\to 0$ it follows that 
$P(\rho) \to 0$ as $\rho\to 0$.
Moreover, by (\ref{hprime}),
\[
\frac{dP}{d\rho} = \frac{dP}{dy} / \frac{d\rho}{dy}
= \frac{1}{3}
\frac{4 g_{3/2}(y) + 3 g_{1/2}(y)}{4 g_{3/2}(y) + 5 g_{1/2}(y)
 + g_{-1/2}(y)} < \frac{1}{3} < 1
\]
for all $0<\rho<\infty$, and also $\frac{dP}{d\rho}>0$,
so that the assumption (A.0) in \cite{Ma} is satisfied.

The assumption (A.1) in  \cite{Ma}, namely that for $\rho \to 0$,
\[
\frac{\rho}{P} \frac{dP}{d\rho} = \gamma + \mathrm{O}(\rho^{\gamma -1})
\]
with some constant $4/3 < \gamma < 2$, is needed only to make sure that
the solutions of the system (\ref{tov1}), (\ref{tov2}), (\ref{eqofstate}) 
have finite radius. Since we know this a priori by the assumption
(\ref{phiass}), we need not check this condition. But using the results
of \cite{RR4} one finds that (A.1) holds with $\gamma=(k+5/2)/(k+3/2)$
which lies in the required interval $]4/3,2[$ iff $-1/2 < k < 3/2$,
and this is precisely the assumption on $k$ in (\ref{phiass}) for $l=0$.

The final condition (A.2) which we need to check is that
the limit
\[
\lim_{\rho \to \infty} \frac{P}{\rho} = \alpha_0 \in ]0,1[
\]
exists, and that there exists some non-decreasing function $\omega$
such that
\[
\left|\frac{P}{\rho} - \alpha_0 \right| \leq \omega\left(\frac{1}{P}\right)
\ \mbox{and}\ 
\int_0^1 \frac{\omega(x)}{x} dx < \infty.
\]
For this it is useful to introduce the functions
\[
h_j(y):= \int_y^1  \phi(\eta) (\eta^2-y^2)^j d\eta,\ y\in [0,1],
\]
so that 
\[
g_j(y) = y^{-(2j+1)} h_j(y),\ y\in ]0,1].
\]
Then
\[
\lim_{\rho \to \infty} \frac{P}{\rho}
= \frac{1}{3} \lim_{y \to 0} \frac{g_{3/2}(y)}{g_{3/2}(y) +  g_{1/2}(y)}
= \frac{1}{3} \lim_{y \to 0} \frac{h_{3/2}(y)}{h_{3/2}(y) + y^2 h_{1/2}(y)}
= \frac{1}{3};
\]
note that $h_{3/2}(0) >0$. Moreover,
\begin{eqnarray*}
\left|\frac{P}{\rho} - \frac{1}{3}\right| 
&=&
\frac{1}{3} \frac{g_{1/2}(y)}{g_{3/2}(y) +  g_{1/2}(y)} \leq 
\frac{1}{3} \frac{g_{1/2}(y)}{g_{3/2}(y)}\\
&=&
\frac{1}{3} \left(\frac{c_0}{3}\right)^{1/2} \frac{g_{1/2}(y)}{(g_{3/2}(y))^{1/2}}
\frac{1}{(c_0 g_{3/2}(y) /3)^{1/2}}\\
&=&
\frac{1}{3} \left(\frac{c_0}{3}\right)^{1/2} \frac{g_{1/2}(y)}{(g_{3/2}(y))^{1/2}}
\frac{1}{\sqrt{P}}.
\end{eqnarray*}  
If we can show that $g_{1/2}(y)\,(g_{3/2}(y))^{-1/2}$ is bounded then the
condition (A.2) holds with $\alpha=1/3$ and $\omega (x)= c \sqrt{x}$, where
$c>0$ is a suitable constant. Now
\[
\frac{g_{1/2}(y)}{(g_{3/2}(y))^{1/2}} =  
\frac{h_{1/2}(y)}{(h_{3/2}(y))^{1/2}},
\]
and the functions $h_{1/2}$ and $h_{3/2}$ are continuous and positive
on $[0,1[$. Hence it remains to show that the above fraction has a finite
limit as $y\to 1$. Let us denote $\varepsilon := 1-y^2 \to 0$ as $y \to 1$.
Then by \cite[(3.18)]{RR4},
\[
\frac{h_{1/2}(y)}{(h_{3/2}(y))^{1/2}}
=
\frac{\frac{1}{2} c_{k,1/2} \varepsilon^{k+3/2} 
+ \mathrm{O}(\varepsilon^{k+3/2+\delta})}
{\left(\frac{1}{2} c_{k,3/2}\varepsilon^{k+5/2}  
+ \mathrm{O}(\varepsilon^{k+5/2+\delta})\right)^{1/2}} \to 0
\]
as $\varepsilon \to 0$, since $k+3/2 > (k+5/2)/2$ which is equivalent to 
$k>-1/2$.  The latter holds by the assumption on $k$ in
(\ref{phiass}), and the proof is complete. \hspace*{\fill} $\Box$ 

\begin{figure}[htbp]
\begin{center}\scalebox{.6}{\includegraphics{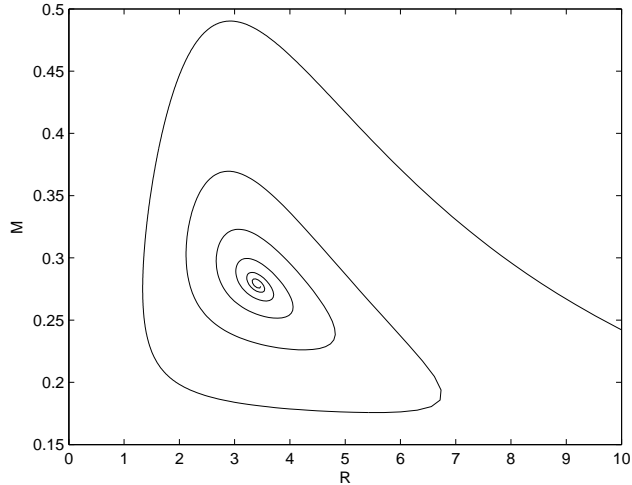}}
\end{center}
\caption{$(R,M)$ spiral for $k=0,\ l=10.5,\ L_0=0,\ 0.01 \leq y(0) \leq 0.99$}
\label{nonisospiral}
\end{figure}

\begin{figure}[htbp]
\begin{center}\scalebox{.6}{\includegraphics{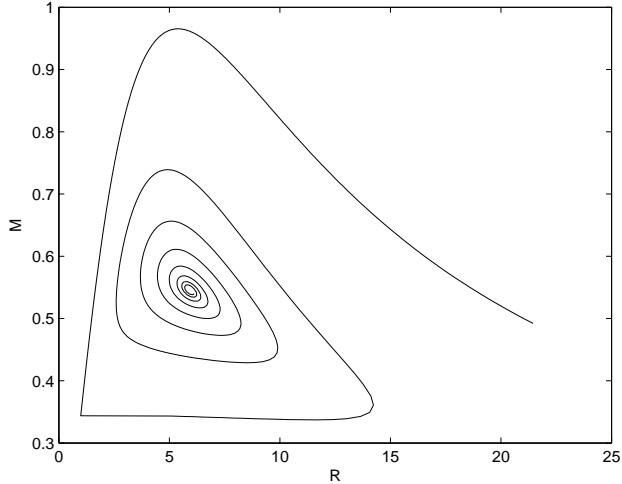}}
\end{center}
\caption{$(R,M)$ spiral for $k=1,\ l=5,\ L_0=2.0,\ 0.01 \leq y(0) \leq 0.99$}
\label{nonisospiral2}
\end{figure}

The theorem above is restricted to isotropic steady states, but
numerically we have so far found no microscopic equation of
state, isotropic or not, which does not lead to a radius-mass
spiral, although for sufficiently anisotropic cases the spiral
becomes rather deformed. Corresponding examples are given 
in Figures~\ref{nonisospiral} and \ref{nonisospiral2}.
  
The sharp corner in Figure~\ref{nonisospiral2} is a genuine 
feature of this curve and can be explained as follows. For
$L_0>0$ the steady state can consist of several shells of matter separated
by vacuum. For $y(0)$ close to 1 only a single shell is present,
but at a certain value of $y(0)$ a second shell appears at a certain
radius which causes the outer radius of the steady state 
to increase discontinuously while the mass remains more or less constant;
recall that $\rho(r) >0$ iff $y(r) \sqrt{1+L_0/r^2} < 1$, and although
$y(r) \sqrt{1+L_0/r^2}$ depends continuously on $y(0)$ the region where
the former condition holds can change discontinuously.     

We conjecture that such radius-mass spirals are a general feature
of one-parameter families of steady states of the spherically
symmetric Einstein-Vlasov system. This is surprising, since it is 
by no means true for the Vlasov-Poisson system. For example,
the  polytropic ansatz
\[
f(r,w,L) = \left(E_0 -  E\right)_+^k L^l
\]
with $k, l>-1$, $k+l>-1/2$, and $k< 3l+7/2$ is known to lead to 
steady states with finite mass and compact support; the particle energy
$E$ is given in terms of the gravitational potential $U(r)$ by
\[
E =  U(r) + \frac{1}{2}(w^2 + L/r^2).
\]
If we keep $k$ and $l$ fixed we obtain a one-parameter family of steady states
which is conveniently parameterized by $y(0)=E_0-U(0)$. 
The resulting equation for $y=E_0-U$ is
\[
\frac{1}{r^2} (r^2 y')' = - 4 \pi c_{k,l} r^{2l} y_+^{k+l+3/2},
\]
where $c_{k,l}$ is some positive constant. One can obtain any solution of this
equation by properly rescaling a fixed, particular one, say, the one with
$y(0)=1$.
Using this fact it is easy to show that if $k-l-3/2\neq 0$
there exists for each prescribed mass $M\in ]0,\infty[$ exactly one steady state 
in the family with that mass, and the relation between radius and mass is
\[
M=R^{(k-l-3/2)/(k+l+1/2)} .
\]
If $k-l-3/2 = 0$ all steady states of the family have the same mass.
In the relativistic case the picture is very different, even for the same type
of ansatz. If the prescribed mass is too large or too small there does not
seem to be a corresponding steady state in the family, while 
for masses in the intermediate range there are in general more than
one steady state of that mass. For the mass $M=M_S$ corresponding 
to the focus of the spiral, cf.~Theorem~\ref{rmspirals}, there are 
infinitely many steady states of that mass, which differ in their radii.

Further interest in these radius-mass spirals comes from the fact that
in astrophysical investigations conclusions about stability of the
steady states in question are often drawn from them via the 
``Poincar\'{e} turning point principle'': If one knows from somewhere
that one steady state on the upper part of the spiral to the right
of its first maximum is stable then it is claimed that all steady states
up to the first maximum of $M$ along the spiral are stable, and that
towards the left of that point stability is lost. We can unfortunately not
claim to understand the arguments by which this principle is supported,
in particular not in the case of an infinite dimensional dynamical system
such as the Einstein-Vlasov system, and we are not aware of any rigorous
mathematical result on stability for this system beyond the
preliminary results in \cite{Wo}. 
However, 
numerically we found in \cite{AnRe1} that along such a one-parameter
family of steady states stability is lost if the so-called binding energy
passes its maximum, and numerically it turns out that this happens at
the same parameter value where $M$ attains its maximum. 
Hence the prediction of the turning point principle 
agrees with the numerical stability analysis in \cite{AnRe1}.

Above we showed analytically that radius-mass spirals are not always
present in the Newtonian case. An example where numerically one 
does find a radius-mass spiral is
the so-called King model where
\[
f(r,w,L) = \left(e^{E_0-E}-1\right)_+.
\]
This model is important in astrophysics. According to the
turning-point principle outlined above there should be a transition
from stable to unstable steady states along the corresponding one-parameter
family, but it has been proven in \cite{GR4} that all steady states
of King type are non-linearly stable; the
analogous result for the so-called relativistic Vlasov-Poisson system 
has been established in \cite{HR}.
It is conceivable that in the non-relativistic case one must plot
other quantities instead of radius and mass to apply the 
turning-point principle, but if one for example plots the radius versus
the total energy for the King model again a spiral arises, which 
in view of \cite{GR4} does not  agree with the prediction of
the turning-point principle.

We conclude this section by emphasizing that the mathematically
rigorous relation---if any---of such radius-mass spirals to stability 
properties of the steady states seems a very interesting and non-trivial
problem in general and for the Einstein-Vlasov system in particular. 
An also very interesting problem is to
prove that radius-mass spirals are present for the Einstein-Vlasov
system in the anisotropic case as well, as our numerical experiments 
seem to indicate.

\end{document}